\documentstyle[12pt]{article}
\title{
DEVELOPMENT OF HIGH PRESSURE XE \\
SCINTILLATION PROPORTIONAL COUNTER \\
FOR EXPERIMENTS IN  \\
"LOW-BACKGROUND" PHYSICS. \\}

\vspace{2cm}

\author{
~\\
~\\
D.Yu. Akimov\footnotemark[1], A.A Burenkov, D.L. Churakov, \\
V.F. Kuzichev, V.L. Morgunov, G.N. Smirnov, V.N. Solovov \\
~\\
~\\
~\\
  \small{Institute for Theoretical and Experimental Physics} \\
  \small{RU-117259  Moscow  Russia} \\         }

\date{}

\begin{document}
\vbox{%
\maketitle}
\newpage

\begin{abstract}

Characteristics of a scintillation proportional counter with WLS fiber 
optics readout is described. The possibility of detection of the 
proportional scintillation signal produced by the single electron of primary 
ionization is shown. The counter can be applied for the experiments in 
"low--background" physics which require a low energy threshold.
\footnotetext[1]{E-mail : akimov\_d@vxitep.itep.ru}
\end{abstract}

\newpage

~
\vspace{4cm}

\section{Detector}

A detector with a very low energy threshold and a significant mass of working 
substance is actual now for the modern experiments in 
"low--background" physics such as a detection of the weakly interacting 
particles of dark matter in the Universe \cite{mosc, cald, absm, cline} and 
measurement of neutrino magnetic moment with the use of artificial neutrino 
source \cite{bern}.

We investigate an original method of readout for gas scintillation proportional 
counter which was not used earlier, although it is based on a combination of 
the known physical phenomena studied in many world laboratories \cite{kuzm}. 
This method gives one the new possibility to utilize 
high pressure gas proportional counters for the experiments mentioned above.

With respect to proportional counter with an electrical signal readout 
a scintillation proportional counter has the following basic advantages: 

-- An equivalent electronic noise is $ 5 - 10$ times lower owing to the use of 
a PMT as a low--noise device. 

-- Theoretical limit of energy resolution is lower. 

-- There is no microphonic noise.

-- A HV circuit and spectrometric electronics are totally decoupled from 
each other.

The detector is being a cylinder volume with a central anode wire 
surrounded with an array of fibers (Fig. 1). 

The electrons originated from ionization of the working medium (Xe) are 
collected on the central anode of the detector, where the process of avalanche 
multiplication (gas gain) takes place. An excitation of Xe molecules with 
subsequent emission of the ultraviolet photons ($\sim$ 170 nm; 
electroluminescence or proportional  scintillation) take place 
simultaneously. The photons are coming on the first--stage converter (or 
wavelength shifter; WLS; p--terphenyl) deposited on the fiber surface and than 
re--emitted in the waverange of a soft ultraviolet. These photons penetrate 
into the fiber core, where they are absorbed by the secondary--stage converter 
(POPOP) and re--emitted again in the blue wavelength range, where a PMT 
possesses a maximum sensitivity and a fiber has a maximum transparency.

The gas gain is supposed to be a few units of $10^3$. With this value the total 
light gain (gas gain with electroluminescence) will be $\sim (0.5 - 1.0) 10^4$ 
UV--photons per one electron of primary ionization (one free electron in the 
detector sensitive volume).

For the light collection with WLS fibers the efficiency can be obtained to 
be $\sim$ 0.006 phe/UV--ph (photoelectrons per one UV-photon)
\cite{pars}. This relatively low value is caused 
by a low 
efficiency of light capture in a fiber ($\sim$0.07) and a low quantum 
efficiency of 
a PMT photocathode ($\sim$0.15). Nevertheless, owing to the use of the dual 
multiplication (gas gain + light gain) near the anode one can obtain a 
unique yield of $\sim$ 400 - 500 phe/keV and therefore, the sensitivity to the 
SINGLE free electron originated in the detector.

The division of the fibers into groups with the PMT pairs 
viewing fibers from opposite ends operating in coincidence and majority 
coincidences between the groups of fibers will allow one to exclude totally the 
noise of photomultipliers and electronics and to lower the energy threshold 
of triggering down to the level corresponding the energy necessary for 
origination of several electrons of primary ionization.

The detector on the basis of high pressure (up to 20 atm) gas scintillation 
proportional counter with a new method of signal readout can be competitive 
with low--temperature bolometers being developed now for dark matter search 
\cite{absm} because it has low energy threshold and can exceed they by mass 
(it can 
be up to a few tens of kilograms). This method also opens the possibility to 
carry out a unique experiment on neutrino magnetic moment search with the use 
of the low--energy antineutrino tritium source \cite{korn}. For this experiment 
a massive  detector having a low energy threshold ($<$ 1 keV) is required.

Xe is chosen as a working medium owing to the following reasons:

-- It has the high cross section of spin--independent interaction (because of 
the large number of nucleon in the Xe nucleus) and also the isotopes with a spin 
1/2 (129Xe; 26.4\%) and 3/2 (131Xe; 21.2\%) in the natural Xe, which take part in 
spin--dependent interaction with dark matter particles.

-- There are no long--life radioactive isotopes.

-- High degree of purification of large amounts from radioactive contamination 
is possible.

-- It has high cross--section of absorption of the low--energy photons 
(E $<$ 20 keV). This fact allows one to reject effectively the characteristic 
emission of the detector body by means of the near--wall active layer of gas 
(with its own cathodes and anodes, as shown in Fig. 1).

\section{ The use of fiber optics for readout of electroluminescence in the Xe 
based detectors}

Photons of primary and secondary scintillation (electroluminescence) in Xe 
have the short wavelength of 170 nm and can be detected by means of an 
ultraviolet photomultiplier or usual PMT with a wavelength shifter (WLS). 
The first use of optical WLS fibers for this aim was described in \cite{pars} 
(see 
above). Fibers was utilized for original solution of a complicate technical 
problem of the readout of electroluminescent signal in a high pressure 
gas detector and obtaining the event coordinates. The WLS fibers was coated 
with the p--terphenyl by means of vacuum deposition. The possibility of the 
use of WLS fibers for detection of an electroluminescent ultraviolet signal 
is considered in the ICARUS collaboration \cite{cline2}. The fibers are 
proposed 
to be located in a LXe and VLPCs possessing a high quantum efficiency 
($\sim$ 80\%) 
to be used for signal readout. Such a combination (fiber and VLPC) 
allows one to achieve the high value $(\sim 10^{-2})$ of photoelectron yield 
per  one UV photon.

We propose to use WLS fibers in a long cylindrical high pressure gas 
proportional counter for detection of the electroluminescence at the central 
anode. Preliminary study of the detection efficiency of fiber was carried 
out with the setup schematically shown in Fig. 2. The 40--cm long scintillation 
fiber with a diameter of core (p--terphenyl + POPOP) of 0.8 mm and a cladding 
diameter of 1 mm was used. One end of the fiber was coupled with the 
central part of the FEU--85 PMT (without optical contact). The PMT current was 
measured.

Ultraviolet photons (170 nm) was emitted from the special UV source on the 
basis of gaseous Xe, radioactive source, and $MgF_2$ window. Since ultraviolet 
light with such a wavelength is totally absorbed in the cladding of the fiber 
two versions of light re--emission before entering the cladding (Fig. 2a and 
Fig. 2b) was tested.
In the case (a) the WLS of the first stage (p--terphenyl) was deposited in 
vacuum on the fiber surface. In the case (b) the WLS was deposited on the 
transparent Mylar film located near the fiber. The photons re--emitted at the 
first stage penetrate into the core where re--emitted again by the secondary 
WLS (POPOP). Since the p--terphenyl coating is mat the process of light 
transportation to the photodetector takes place in the first case only at 
the core--cladding interface while in the second case, at both surfaces. 
Taking this into account one can explain the behavior of the 
distance--efficiency curves for both cases shown in Fig. 3 in relative units. 
The fiber uncoated with a WLS possesses a significant nonuniformity 
(squares) since the outer surface of it is of insufficient quality.

\section{ Energy resolution and light gain of 
electroluminescent detectors}

The use of electroluminescence (proportional scintillation) attracts an 
attention of elaborators of gas proportional detectors owing the possibility 
to obtain a significant internal light multiplication resulting in   
better signal--noise ratio than for electrical readout of signal (with gas 
multiplication) owing to the use of a PMT as a low--noise device. In addition, a  
proportional scintillation counter also shows better energy resolution in 
the multiplication mode with low or unit gas gain \cite{palm}. 

The energy resolution of a scintillation proportional counter operating in 
the mode of gas multiplication is defined with a fluctuation of the following 
values: the number of electrons of primary ionization, the gas gain, and the 
number of photoelectrons:

\begin{equation} %1
(\frac{\sigma_{E}}{E})^2 = \frac{F}{N_e} + \frac{f}{N_e} + \frac{1}{N_{phe}}
\end{equation}

where:

$F$ is Fano factor, $f$ is a factor describing a fluctuation of the gas 
gain \cite{alkh}, $N_e$ is a number of electrons of primary ionization, 
$N_{phe}$ is a number of photoelectrons. 

Taking into account that $N_{phe} = k N_e$ 
the energy resolution can be written as following:

\begin{equation} %2
(\frac{\sigma_{E}}{E})^2 = \frac{1}{N_e}(F + f + \frac{1}{k})
\end{equation}

 where $k = k_{L}\varepsilon Q$ is a 
coefficient of 
transformation (that is the number of photoelectrons per one electron of 
primary ionization), $k_L$ is a light gain, $\varepsilon$ is an efficiency 
of light 
collection to the photocathode, Q is a photocathode quantum efficiency.
The value of Fano--factor for Xe is from 0.15 to 0.17 \cite{lima}, the value 
of f is about 0.5 and $0.7 - 0.8$ for the value of gas gain of $\sim 10$ 
and $>$ $10^2$, respectively \cite{alkh}. Therefore, it is clear that at 
$k$ $>$ 5 the energy resolution 
is defined mainly with the gas gain fluctuation. To improve the energy 
resolution of a scintillation proportional counter one have to reduce the 
gas gain down to the minimal possible value keeping $k$ constant.
The last is achieved by increasing the light collection efficiency $\varepsilon$
and quantum efficiency of a photodetector $Q$. Another way is optimization 
of the wire diameter to obtain a maximum contribution of the 
electroluminescence taking place outside the avalanche region to the total 
light gain. 

The number of photons created by the single electron per unit 
of length is calculated according to the semi--empirical formula:

\begin{equation} %3
 \frac{dW_{ph}}{dx} = 70 (E/p - 0.8) p [cm^{-1}]
\end{equation}

where:

 $E$ [kV/cm] is an electric field strength, $p$ [atm] is a pressure.

Then the light gain for a gas counter of cylindrical geometry (without gas 
gain) can be calculated as following:

\begin{equation} %4
k_L = \frac{70 U_a}{\ln(R_c / R_a)} \int \limits_{R_a}^{r_{thr}} \frac{dr}{r}
-56 p  \int \limits_{R_a}^{r_{thr}} dr
\end{equation}

where:

 $U_a$ [kV] is an anode potential, $R_c$, $R_a$ [cm] are cathode and anode 
diameters , respectively, $r_{thr}$ is obtained from the relation:

\begin{equation} %5
\frac{U_a}{r_{thr}\ln (R_c/R_a)} > 0.8
\end{equation}

\vspace{0.5cm}

\section{ Experimental tests with the prototype}

The prototype described in \cite{akim} and shown in Fig. 4 was built to study 
characteristics of a scintillation proportional counter with fiber optical 
readout. A fiducial length and an inner diameter of the counter are 25 cm 
and 2.2 cm, respectively. The fibers with a total number of 72 are located 
near the inner surface of the cylindrical wall. The ends of the fibers are 
brought together in a wisp near the glass window having a diameter 1.5 cm. A 
PMT of FEU--85 type is placed behind this window outside the Xe volume. The 
70 {$\mu$}m stainless steel cathode wires form a right octagon and located at a 
distance of 1.1 cm from the cylinder axis. The cathodes have a potential equal 
to that of the counter wall. The gilt tungsten anode wire with a 
diameter of 50 {$\mu$}m is stretched along the cylinder axis. A radioactive 
source $~^{241}Am$ is put on the inner wall surface at a distance of 5 cm from 
the end of the counter. 

The prototype is designed for the gas pressure of up to 8 atm. The value of 
pressure is restricted by a strength of the optical window used. Gas 
filling of the counter was performed through an "Oxisorb" purifier. 
Since a constant circulation and purification of the gas was not provided 
gas refreshing in the counter was performed by means of refreezing it 
back to the storage cylinder and passing it again through the "Oxisorb".

The scheme of electronics is shown in Fig. 5. A readout of an electrical signal 
from the wire was performed by means of charge--sensitive preamplifier in 
parallel with electroluminescent readout. The PMT signal passed through 
the circuit of discrimination and selection which rejects the pulses with a 
duration T $<$ 250 ns is used as a trigger. Both signals are digitized by means 
of charge--to--digit converters. 
A light pulse from the light emitting diode (LED) coupled with the photocathode 
of PMT through an optical fiber is used for calibration.

\section{ Experimental results}

The pulse height spectra measured with $~^{241}Am$ radioactive source at a 
pressure of 2 atm and $U_a$ = 2.9 kV for the electroluminescent 
and electrical readout are shown in Fig. 6 and Fig. 7, 
respectively. It is seen from these spectra 
that both channels are totally equivalent each other, the energy resolution 
%                              ^^^^^^^^^^
%????? Why do we need the fibers if we have the EQUIVALENT resolution????
%
for each case is the same (13\% (FWHM) for 13.9 keV and 12\% for 
17.8 keV). The yield of photoelectrons is obtained to be 450 phe/keV, and 
thus, the $k$ = 11 photoelectrons per one electron of primary ionization. 
According to (2) for such value of $k$ the resolution to be 10\% at 13.9 keV. 
The worse value of energy resolution can be explained by possible 
nonuniformity of the wire diameter. Fig. 8 shows the pulse height spectrum 
obtained for the electroluminescence channel at a pressure of 8 atm, the 
energy resolution is 13\% at 13.9 keV. A low--energy part of the spectrum ($<$ 
15 keV) specially measured at 8--atm pressure is shown in Fig. 9. The 
maximum value of a photoelectron yield achieved at a pressure of 8 atm is 
230 ph e/keV. It is seen from Fig. 9 that with the triggering mode used 
(single PMT + time selection of pulses) the energy threshold can be set at 
a level of 0.2 keV. 

To demonstrate the possibility to lower the energy threshold down to the 
level corresponding the energy necessary for origination of several primary 
electrons we have shown experimentally the detection of 
the single electroluminescent pulse, which corresponds to the process of 
avalanche multiplication caused by a single electron. For this aim we used the 
feedback pulses coming after the main pulse. In our case at a pressure 
of 2 atm  and $k$=7 the main pulse was accompanied with a sequence of feedback 
pulses having a period equal to the electron drift time from the cathode to 
the anode ($\sim$ 8 {$\mu$}s) and a ratio of heights of neighboring pulses of 
1:10. 
Therefore, the pulse appearing after the main pulse at the time corresponding, 
for example, to the 6--th feedback pulse must be knowingly a single--electron 
one. Fig. 10 shows the pulse height spectrum (2) for such signals. The signals 
with a total duration of more than 120 ns was selected. This value is higher 
than that of the noise (single--photoelectron) signals. The pulse height 
spectrum of the signals from the PMT operated in a single--photoelectron mode 
and triggered by the LED signal is shown in the same figure (1). The average 
number of photoelectrons in the single--electron electroluminescent pulse is 
obtained to be $7 \pm 1$.

Fig. 11 shows the light gain and gas gain as a function of applied voltage for 
pressures 2, 4, and 8 atm. The experimental data are obtained with {$\alpha$} 
and {$\gamma$}--peaks at 2 and 4 atm while at 8 atm, only with 
{$\gamma$}--peaks 
since at this pressure {$\alpha$}--particles don't reach the sensitive volume.

The value of light collection efficiency {$\varepsilon$}, which is necessary 
for 
obtaining the value of light gain, was estimated from fit with the formula 
(4) of the experimental points measured at such values of applied 
voltage $U_a$ when the gas gain is low ($<$ 5) or zero. 
It is obtained to be $(0.6 \pm 0.1) 10^{-3}$ ph e/photon. Note that this 
value can be increased really by a factor of $3 - 4$ by means of more compact 
disposition of fibers, well matching of spectral characteristics 
of the first and second stage wavelength shifters and photomultiplier, 
individual selection of PMTs with a high quantum efficiency, and readout of 
light signal from both ends of fibers.

The experimental data demonstrate that both ligh gain and gas gain as a 
function of the anode voltage have the same exponential behavior at the tail 
(gas gain $> 10^3$). 
It is the evidence that the electrons born in the avalanche give the 
main contribution to the light gain in spite of the fact that their path 
length is very short (about of several tens of {$\mu$}m). This fact accounts for 
an equal energy resolution obtained both for light and charge readout (see 
Fig. 6 and 7) because the photoelectron statistics is defined mainly by 
the statistics of electrons in the avalanche. 

\section{ Conclusion}

It is shown that with the method of signal readout by means of WLS optical 
fibers the value of $k$ $\sim 5 - 10$ (photoelectrons per one primary electron)  
can be obtained for a high pressure Xe gas proportional counter with a 
cylindrical geometry and having significant sizes. Thus, the efficiency of 
light collection is sufficient for the photoelectron statistics to be defined 
mainly by the statistics of electrons in the avalanche. 

The value of photoelectron yield of $>$ 200 phe/keV is obtained for the pressure 
of up to 8 atm. This fact allows one to obtain for such detectors a low energy 
threshold ($<$ 1 keV) owing to a low noise level of photomultiplier.

For the energy resolution of the scintillation proportional counter to be 
better than that of a proportional counter with electrical readout it is 
necessary to enhance the contribution to the light gain of the 
electroluminescence taking place before the beginning of avalanche 
multiplication process. This can be achieved, as it is seen from Fig. 11, by 
means of reduction $U_a$ (with making ${\varepsilon}$ higher) and increasing a 
pressure.

We would like to thank M. Danilov for a great interest.

\section*{FIGURE CAPTIONS}

Fig. 1. Layout of electrodes and fibers inside the cylindrical body of the 
counter. 1 -- counter body, 2 -- WSL fibers, 3 -- wire anodes, 4 -- wire 
cathodes, 5 -- central anode.\\
\\

Fig. 2. Scheme of the WLS fiber tests. The wavelength shifter is deposited 
on the fiber (a); on the Mylar film (b). 1 -- core, 2 -- cladding, 3 -- 
wavelength shifter, 4 -- light collimator, 5 -- UV source, 6 -- PMT, 7 -- 
current meter.\\
\\

Fig. 3. Light collection efficiency of the fiber irradiated by 170 nm light 
versus the distance from PMT. Triangles and squares are corresponded to the 
cases (a) and (b) of Fig. 2, respectively. \\
\\

Fig. 4. Schematic diagram of the prototype. 1 -- fibers, 2 -- cathode wires, 
3 -- anode wire, 4 -- glass window, 5 -- $~^{241}Am$ source. \\
\\

Fig. 5. The scheme of electronics. G1, G2 -- fine tuning generators, LED -- 
light emitting diode, INV -- invertor, FA -- fast amplifier, CSP -- charge 
-- sensitive preamplifier, DL -- cable delay line, D/S -- 
discrimination/selection unit, M -- the monostable, AC -- anticoincidence 
circuit, D -- delay ($T_{dr}$), ADC1, ADC2 -- charge--digital converters.\\
\\

Fig. 6. Pulse height spectrum obtained with $~^{241}Am$ source via 
electroluminescence channel; 
P = 2 atm, $U_a$ = 2.9 kV. 1 -- pedestal, 
2 -- $L_{\alpha} \; \; Np \; \; (13.9 keV); $
3 -- $L_{\beta} \; \;Np \; \;(17.8 keV); $
4 -- $K_{\alpha} \; \;Xe \; \;(29.8 keV), $
59.6 keV -- $K_{\alpha} \; \;Xe\; 
\;(29.8 keV).$ \\
\\

Fig. 7. Pulse height spectrum obtained with $~^{241}Am$ source via 
charge--collection channel at P = 2 atm, $U_a$ = 2.9 kV. 1 -- pedestal, 
2 -- $L_{\alpha} \; \;Np \; \;(13.9 keV); $
3 -- $L_{\beta} \; \;Np \; \;(17.8 keV); $
4 -- $K_{\alpha} \; \;Xe \; \;(29.8 keV), $
59.6 keV -- $K_{\alpha} \; \;Xe\; 
\;(29.8 keV).$ \\
\\

Fig. 8. Pulse height spectrum obtained with $~^{241}Am$ source via 
electroluminescence channel at P = 8 atm, $U_a$ = 5.5 kV. 1 -- pedestal, 
2 -- $L_{\alpha} \; \;Np \; \;(13.9 keV); $
3 -- $L_{\beta} \; \;Np \; \;(17.8 keV); $
4 -- $K_{\alpha} \; \;Xe \; \;(29.8 keV), $
59.6 keV -- $K_{\alpha} \; \;Xe\; 
\;(29.8 keV).$ \\
\\

Fig. 9. Low-energy part ($<$ 15 keV) of the pulse height spectrum measured at 
8--atm pressure. \\
\\

Fig. 10. Pulse height spectra measured in a single--electron mode (2; $U_a$ = 
2.9 kV, P = 2 atm) and in a single--photoelectron mode (1; $U_a$ = 0). \\
\\

Fig. 11. Light gain and gas gain as a function of applied voltage at a 
pressure of 2, 4, 8 atm. LG is light gain, GG is gas gain, curves -- fit 
with formula (4).

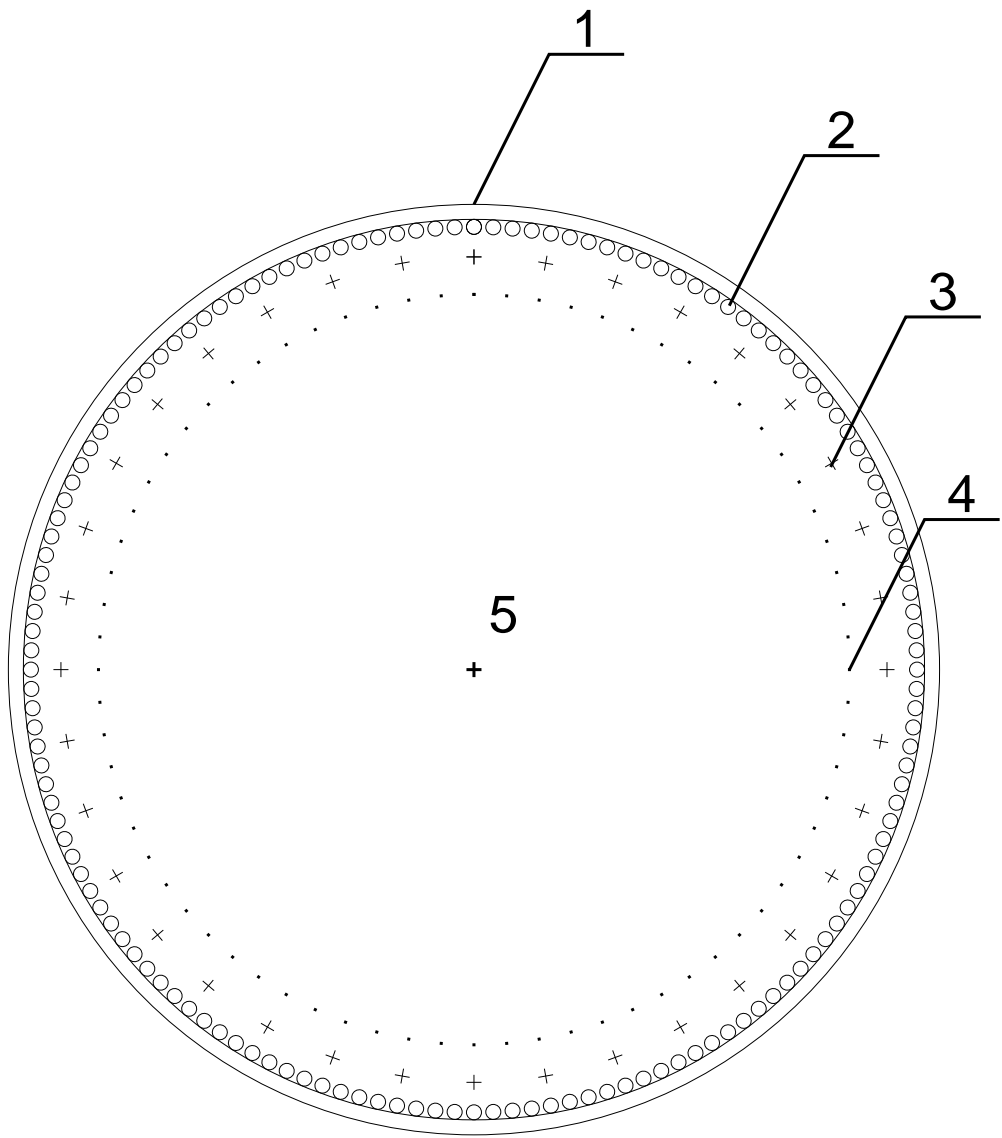
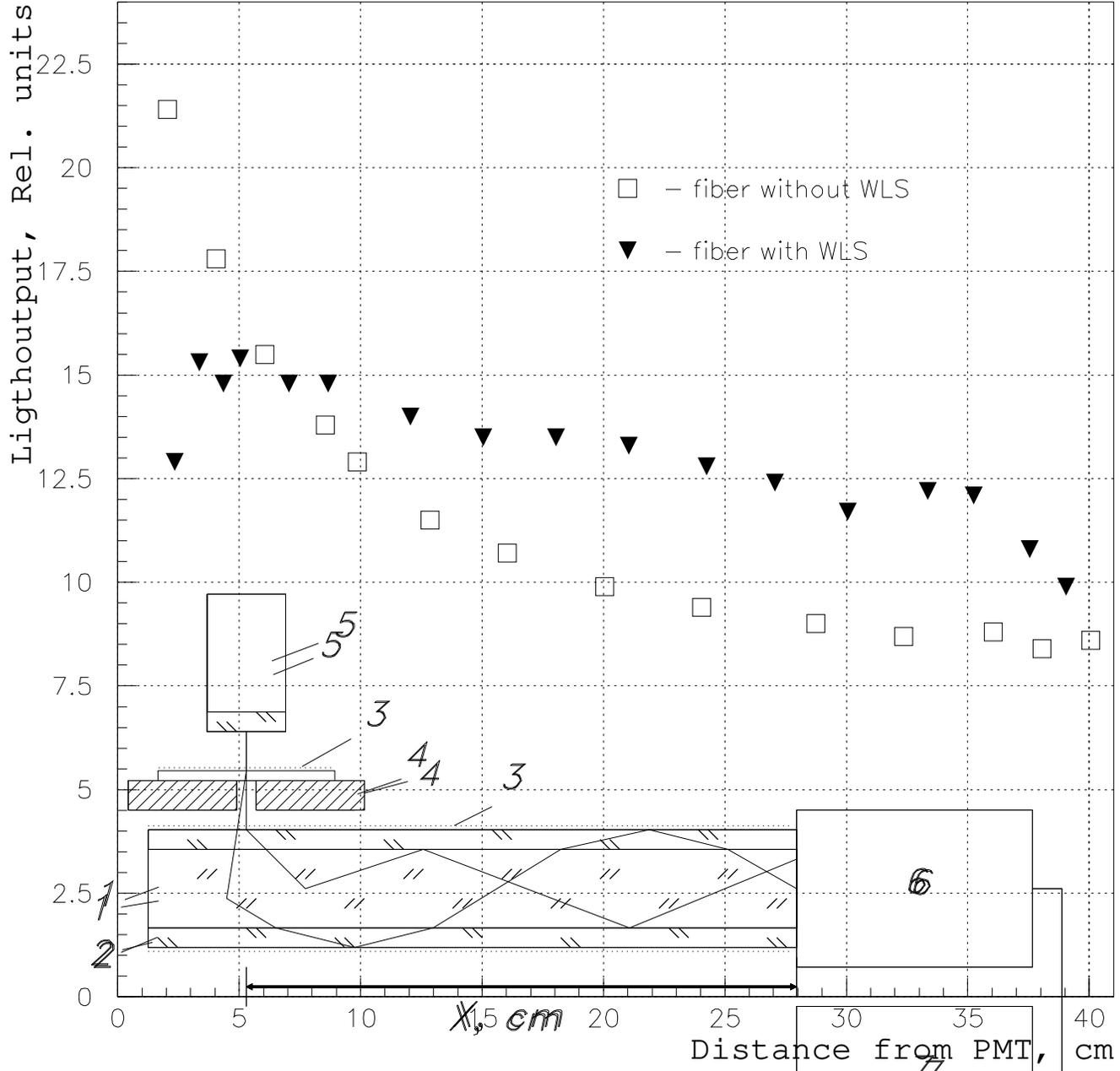
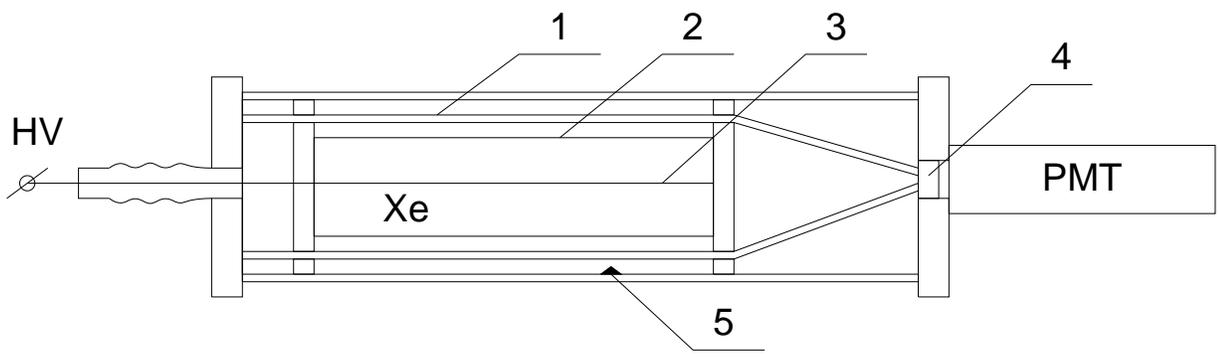
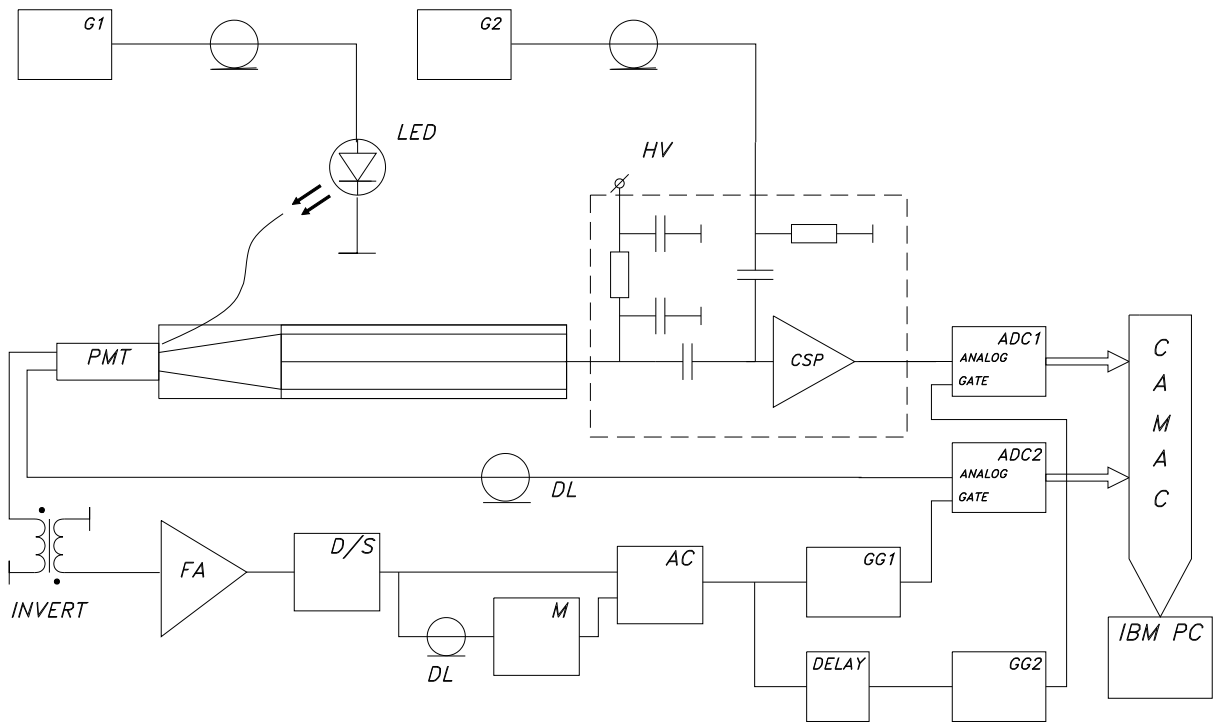
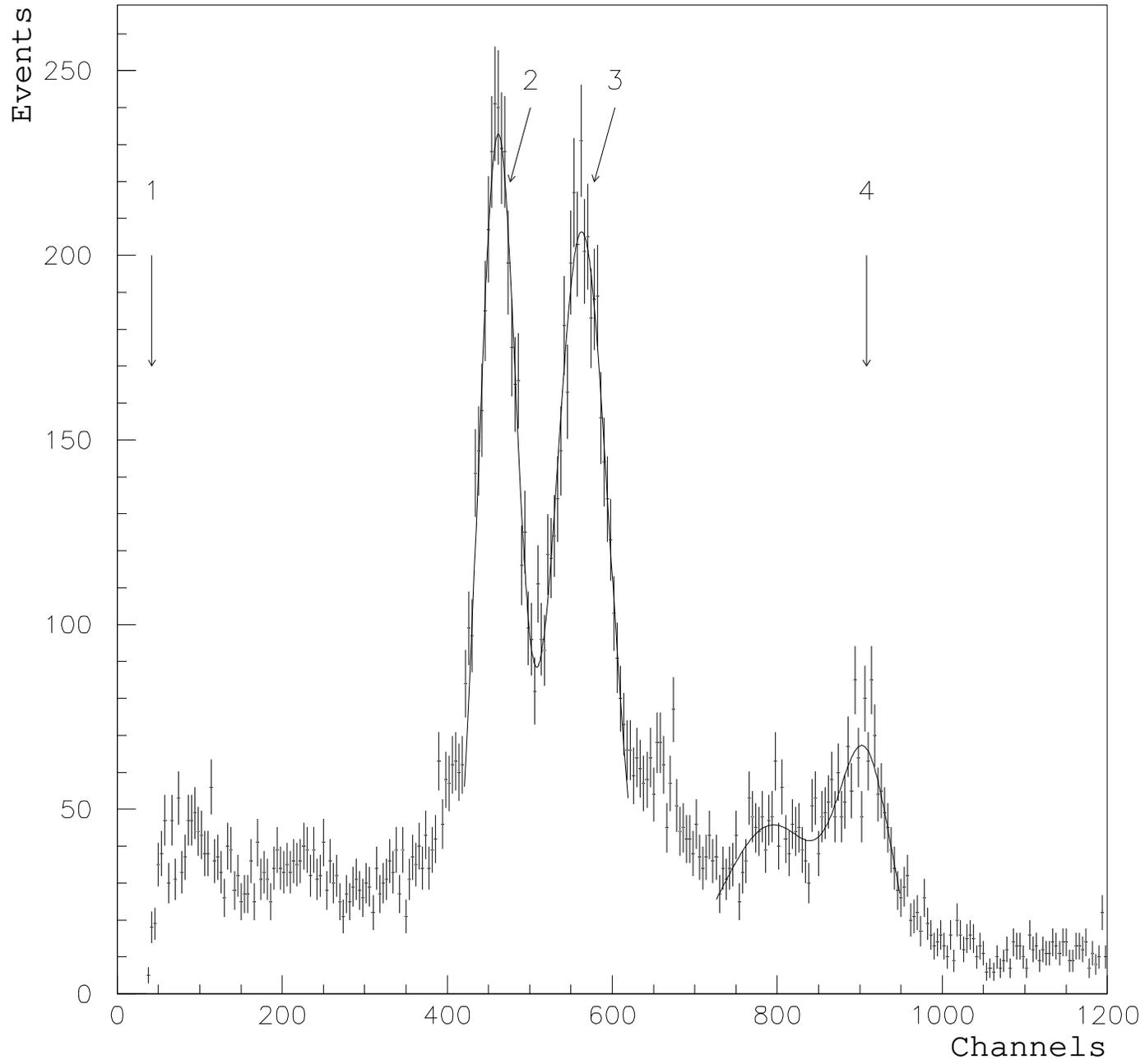
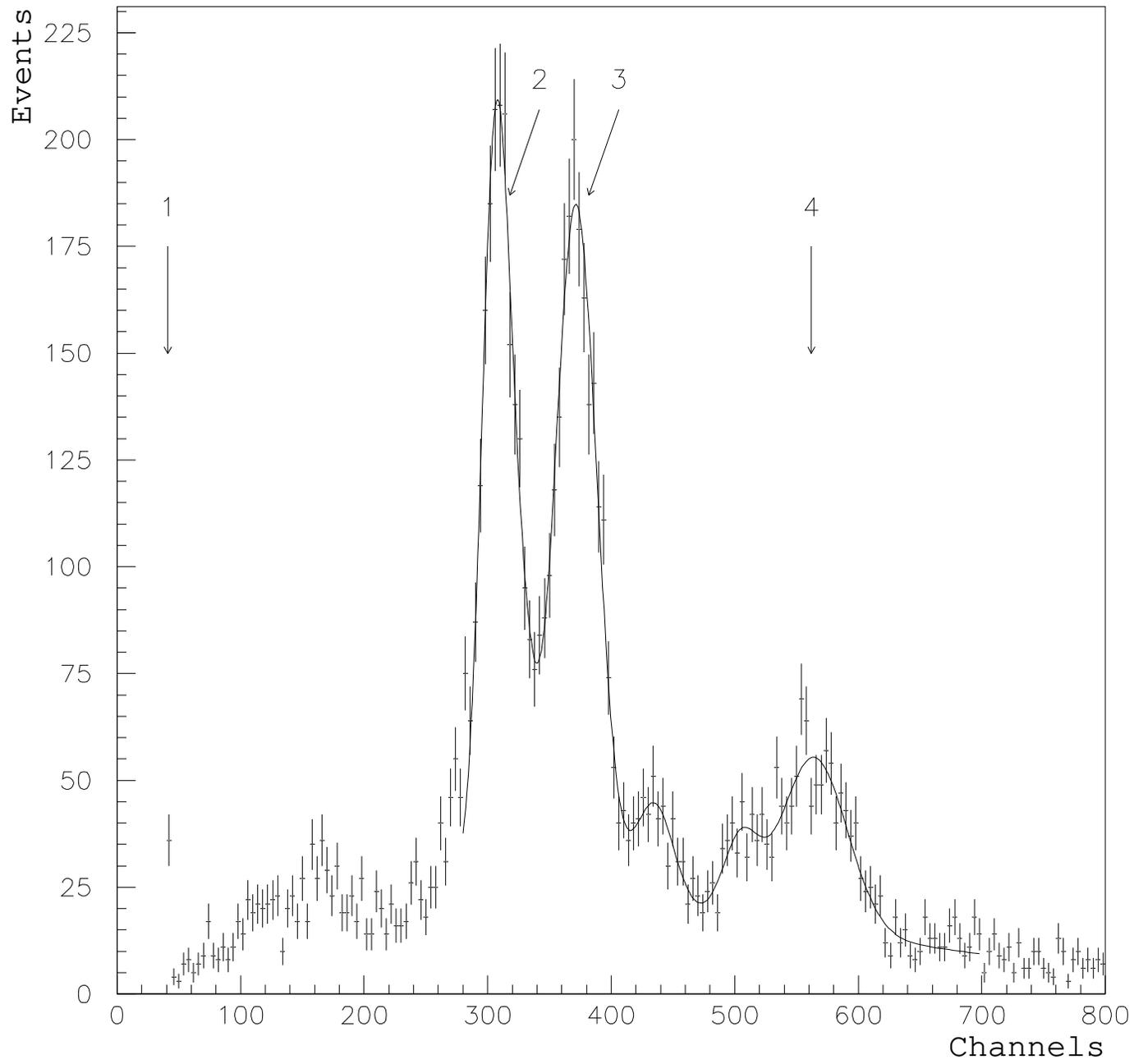
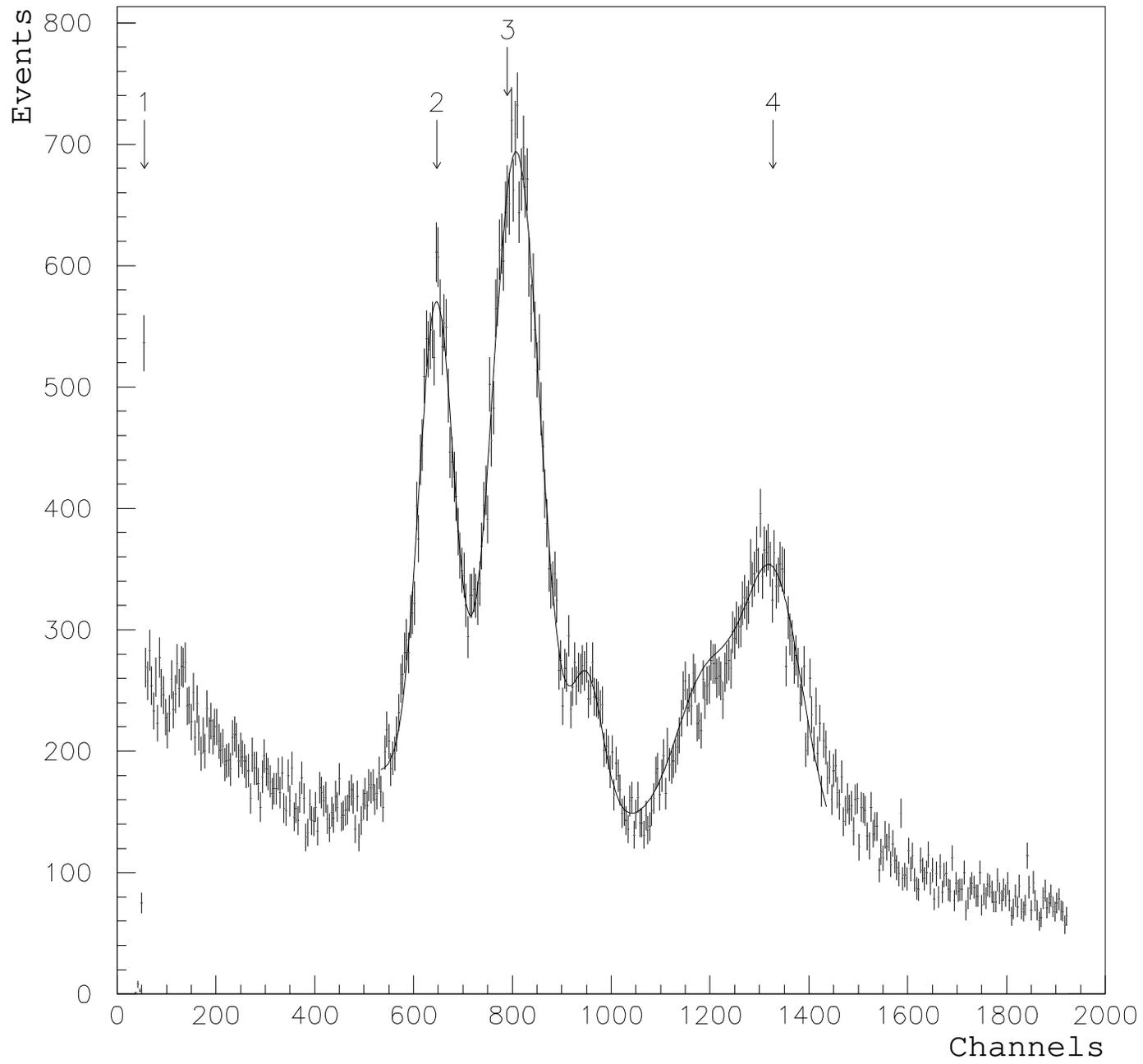
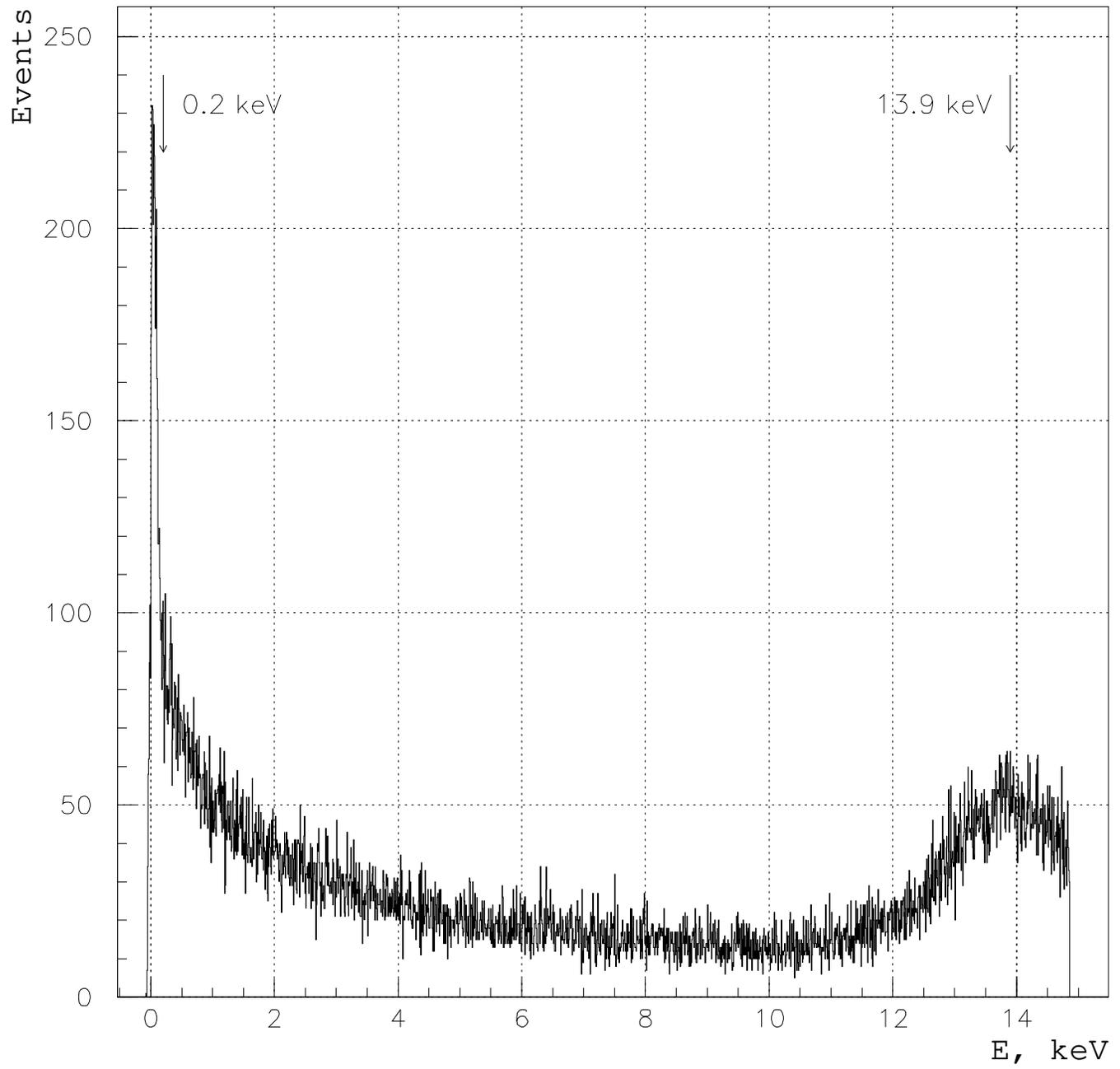
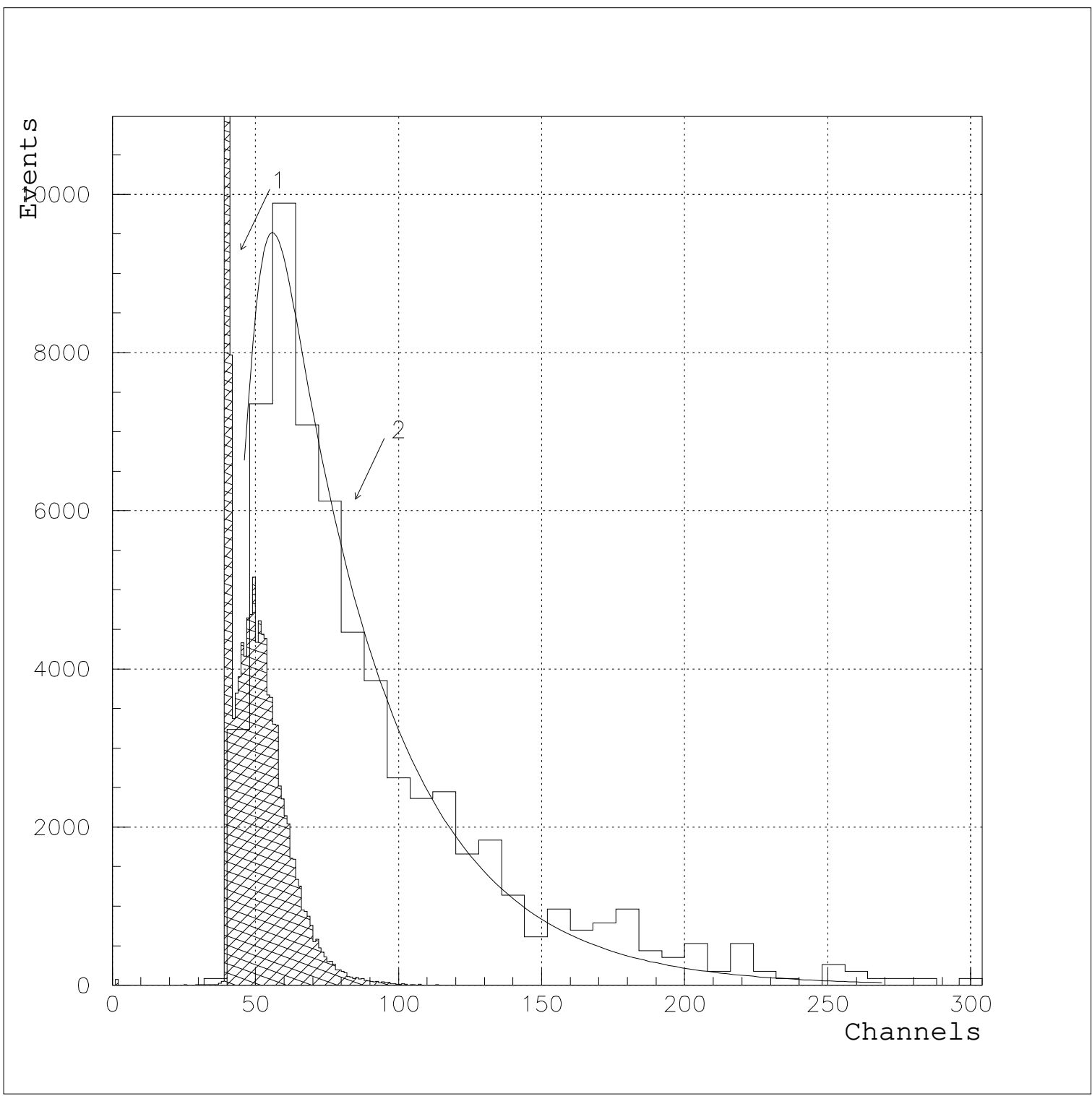
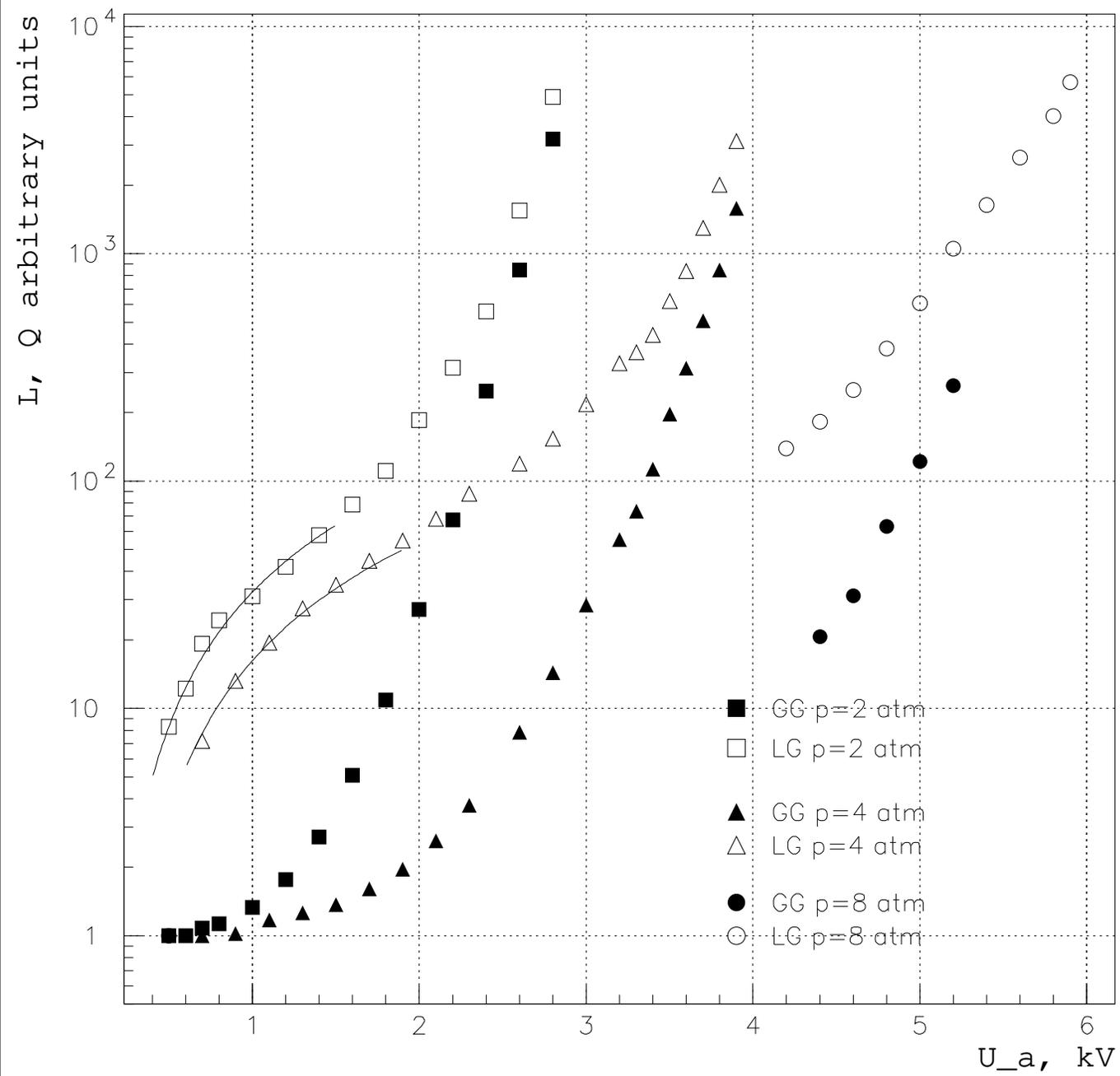
\end{document}